\RequirePackage{tikz} 
\documentclass[doublecol]{epl2}
\usepackage{amssymb}

\newif\ifdraft % False by default?
\ifdraft
  \usepackage{eso-pic}
  \makeatletter
  \AddToShipoutPicture{%
  \setlength{\@tempdimb}{.5\paperwidth}%
  \setlength{\@tempdimc}{.5\paperheight}%
  \setlength{\unitlength}{1pt}%
  \put(\strip@pt\@tempdimb,\strip@pt\@tempdimc){%
      \makebox(-550,0){\rotatebox{90}{\textcolor[gray]{0.70}%
         {\Large \textsf{Draft of \today}}}}
    }%
  }
  \makeatother

  \usepackage[normalem]{ulem} % either use this (simple) or
  \newcommand{\rednote}[1]{\textcolor{red}{#1}}
  \newcommand{\redfootnote}[1]{\footnote{\textcolor{red}{#1}}}

\else  
  \newcommand{\sout}[1]{\iffalse #1 \fi}
  \newcommand{\rednote}[1]{\iffalse #1 \fi}
  \newcommand{\redfootnote}[1]{\iffalse #1 \fi}
\fi

\usepackage{paralist}

\usepackage{graphicx}
\graphicspath{ {figures/}
             }
             
\newcommand{\tikzLabelMod}[3]{
  \begin{tikzpicture}[font=\sffamily]%\bfseries\small]%Helvetika] 
        \node[anchor=south west,inner sep=0] (image) at (0,0){#3};
            \begin{scope}[x={(image.south east)},y={(image.north west)}]
              \draw (0.01,{#2}) node {\ulcase{#1}};
            \end{scope}
        \end{tikzpicture}
    }

\newcommand{\ulcase}{\lowercase} %% TODO
\newcommand{\subfig}[1]{\ulcase{#1}}
\newcommand{\subcap}[1]{(\ulcase{#1})}

\usepackage[binary-units=true]{siunitx} 
\usepackage{xcolor}
\usepackage[colorlinks=true,urlcolor=blue]{hyperref}
\usepackage{breakurl}

\title{Long-range hydrodynamic effect due to a single vesicle in linear flow}

\author{Eldad Afik\inst{1} \and Antonio Lamura\inst{2} \and Victor Steinberg\inst{1}}

\institute{
 \inst{1} Department of Physics of Complex Systems, Weizmann Institute of
          Science, Rehovot, 76100 Israel \\
 \inst{2} Istituto Applicazioni Calcolo, CNR, Via Amendola 122/D, 70126 Bari,
          Italy, EU
        } 
\date{\today}

\abstract{Vesicles are involved in a vast variety of transport processes in living
organisms. Additionally, they serve as a model for the dynamics of
cell suspensions.
Predicting the rheological properties of their suspensions is still an open
question, as even the interaction of pairs is yet to be fully understood. 
Here we analyse the effect of a single vesicle, undergoing tank-treading
motion, on its surrounding shear flow by studying the induced
disturbance field $\delta \vec{V}$, the difference between the velocity field
in its presence and absence.  The comparison
between experiments and numerical simulations reveals an
impressive agreement. 
Tracking ridges in the disturbance field magnitude landscape, we identify the
principal directions along which the velocity difference field is analysed
in the vesicle vicinity.
The disturbance magnitude is found to be significant up
to about 4 vesicles radii and can be described by a
power law decay with the distance $d$ from the vesicle $ \| \delta \vec{V} \| \propto d^{-3/2}$.
This is consistent with previous experimental results on the separation
distance between two interacting vesicles under similar conditions, for which their
dynamics is altered.
This is an indication of vesicles long-range effect via the disturbance
field and calls for the proper incorporation of long-range hydrodynamic interactions when
attempting to derive rheological properties of vesicle suspensions.
}

\pacs{87.16.D-}{Membranes, bilayers, and vesicles}
\pacs{83.50.-v}{Deformation and flow}
\pacs{47.63.-b}{Biological fluid dynamics}

\begin{document}
\maketitle

\section{Introduction}

 A long-range hydrodynamic interaction between objects immersed  in a
fluid is recognized as one of the main ingredients contributed into the
effective viscosity of suspensions of meso-scale particles. 
The nonlocal nature of the hydrodynamic interaction is a key physical factor
that makes this problem  so  rich,  but  at  the  same  time  so  difficult.
The velocity distribution in the fluid near an immersed object is affected by
the presence of another object at a finite distance. It results in modification
of the fluid stresses at the surface of the first object and correspondingly of
its translational and rotational motion. The effect critically depends on the
nature and elastic properties of the immersed objects.
 
 In this respect, central questions are what is the
characteristic spatial scale of a long-range hydrodynamic interaction and
whether and how does it depend on the nature and elastic properties of objects
immersed in a fluid? In the case of solid spherical particles in a fluid
subjected to a linear flow field, this problem was first addressed theoretically
in Ref. \cite{Batchelor1972a} and then studied in many papers both theoretically and
numerically (see e.g. \cite{Morris2009,Sangani2011} and references therein). In the case
of vesicles, the long-range hydrodynamic vesicle pair interaction and the
corresponding spatial scale was investigated recently both experimentally
\cite{Kantsler2008,Levant2012} as well as numerically \cite{Lamura2013}. In this paper we
address the problem directly related to the long-range hydrodynamic interaction
of a pair of vesicles in a planar linear flow, namely: How strong are 
perturbations of a linear velocity field caused by a vesicle and how far away
from it are they observed? What is scaling of the perturbations of the linear
velocity field with the distance from the vesicle? We explore these questions
both experimentally and numerically and compare the results.

A unilamellar liposome vesicle is a drop of fluid encapsulated in a
lipid bi-layer membrane, suspended in either the same or different fluid. The
impermeability and the inextensibility of the lipid membrane dictate
conservation of both volume and surface area of the vesicle. Vesicles undergo a
tank-treading motion in a planar linear flow at internal-to-external viscosity
contrast $\lambda = \eta_{in} / \eta_{out} = 1$, vorticity-to-strain ratio
$\omega/s<2$, and excess area $\Delta = S/R_0^2 -4\pi <1$ \cite{Deschamps2009}.
Here $ R_0 = \left( V \big/ \frac{4\pi}{3} \right)^{^1\!/_3}$ is  the vesicle effective
radius, $V$ is the vesicle volume, and $S$ is the vesicle surface area;
$\omega$ and $s$ are the vorticity and strain rates, defined following
Ref.~\cite{Lebedev2007}.
At a fixed viscosity contrast, the inclination angle $\theta$ decreases with 
increasing $\Delta$ and $\omega/s$ \cite{Lebedev2007,Kantsler2005,Kantsler2006,Zabusky2011}.

\section{Experimental, numerical and data analysis methods}

The details of the experimental methods were presented in Ref.
\cite{Levant2012}. Below we provide a brief overview of the
velocity field measurements in the presence of a vesicle, which are in the
focus in this work. 
The experiments were conducted in a microfluidic four-roll mill apparatus
\cite{lee2007,Deschamps2009}.
The key component of this apparatus is the dynamical trap, which allows long
observation times compared to the vesicles orbit period in the flow. The flow
was driven by gravity and the control parameter $\omega/s$ was set via setting
the pressure drop. 
Time-lapse velocity fields were inferred by means of 2D micro
Particle Image Velocimetry (\si{\micro}PIV) at the mid-plane of the dynamical
trap. The aqueous vesicles suspension was seeded with fluorescent
particles, allowing imaging of the flow in the trap.  The images were then
filtered by a Laplace filter using Gaussian second derivatives \cite{SciPy},
and processed using Gpiv \cite{gpiv} at interrogation windows of $32\times32$
pixels (corresponding to about $8.6\times8.6$~\si{\um\squared}) with $50\%$
overlap. \cite{Levant2012}

The 2D numerical model used here was introduced in Ref. \cite{Lamura2013}. 
A single vesicle with membrane bending rigidity $\kappa$, initial length $L_0$,
initial area $A_0$, and $R_0 = L_0/2\pi$ being the vesicle effective radius, is studied in a
wall-bounded shear flow with a shear rate $\dot{\gamma} = \omega/2 = s/2$, 
which is a specific case of a planar linear flow.
Periodic boundary conditions are used in the $x$ direction; 
the upper and the lower walls slide along this direction at velocities
$\pm u_{wall}$, respectively, resulting in a linear flow profile 
$(V_x, V_y) = (\dot{\gamma}y,0)$, with shear rate $\dot{\gamma} = 2u_{wall}/H_y$,
$H_y$ being the flow cell width.
The system size is $H_x\times H_y =18.85R_0\times 8.12R_0$.
The steady velocity field without a vesicle is taken in a frame of size $L_x\times L_y =
10.47R_0\times H_y$ centred at $(H_x/2,H_y/2)$. When the vesicle is embedded
in the solvent, the steady velocity pattern is recorded in a frame of size
$L_x \times L_y$ centred at $(x_{cm},H_y/2)$, where $(x_{cm}, y_{cm})$ denotes the
position of the vesicle centre of mass. 
The bending rigidity is set to $\kappa=6.58k_B TR_0$, where $k_B T$ is the
thermal energy. This value of $\kappa$ gives rise to a similar amplitude of
undulation modes as for lipid bilayer membranes in 3D (where $\kappa\simeq
10k_BT$). The reduced shear rate is $\dot{\gamma}^* = \dot{\gamma}\tau_c=3$,
where $\tau_c = \eta_{out}R_0^3/\kappa$ is the relaxation time of the vesicle,
so that the Reynolds number is $Re = 0.08$ allowing to neglect inertial
effects. The viscosity contrast is set to $\lambda = 1$.

 \begin{figure*}[tb]
  %\onefigure[width=16cm]{velocityfieldexpsim.eps}}
  \begin{center}
  \begin{tikzpicture}[font=\sffamily]%\bfseries\small]%Helvetika] 
        \node[anchor=south west,inner sep=0] (image) at (0,0)%
            {\includegraphics[width=0.47\textwidth,trim=0 6.5mm 10mm 2.5mm,clip]%
                                                        {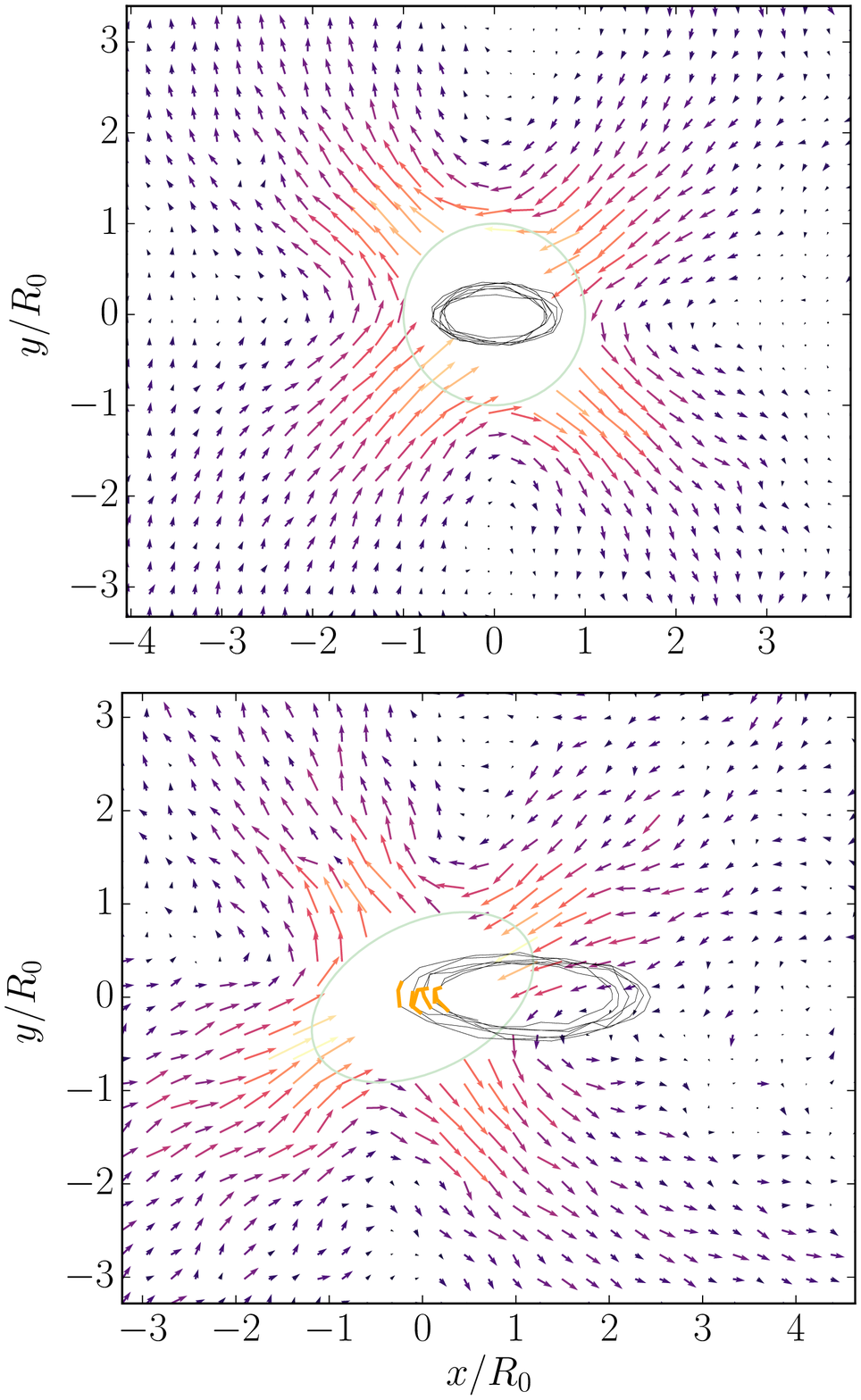}%
             \includegraphics[width=0.47\textwidth,trim=10mm 6.5mm 0 2.5mm,clip]%
                                                        {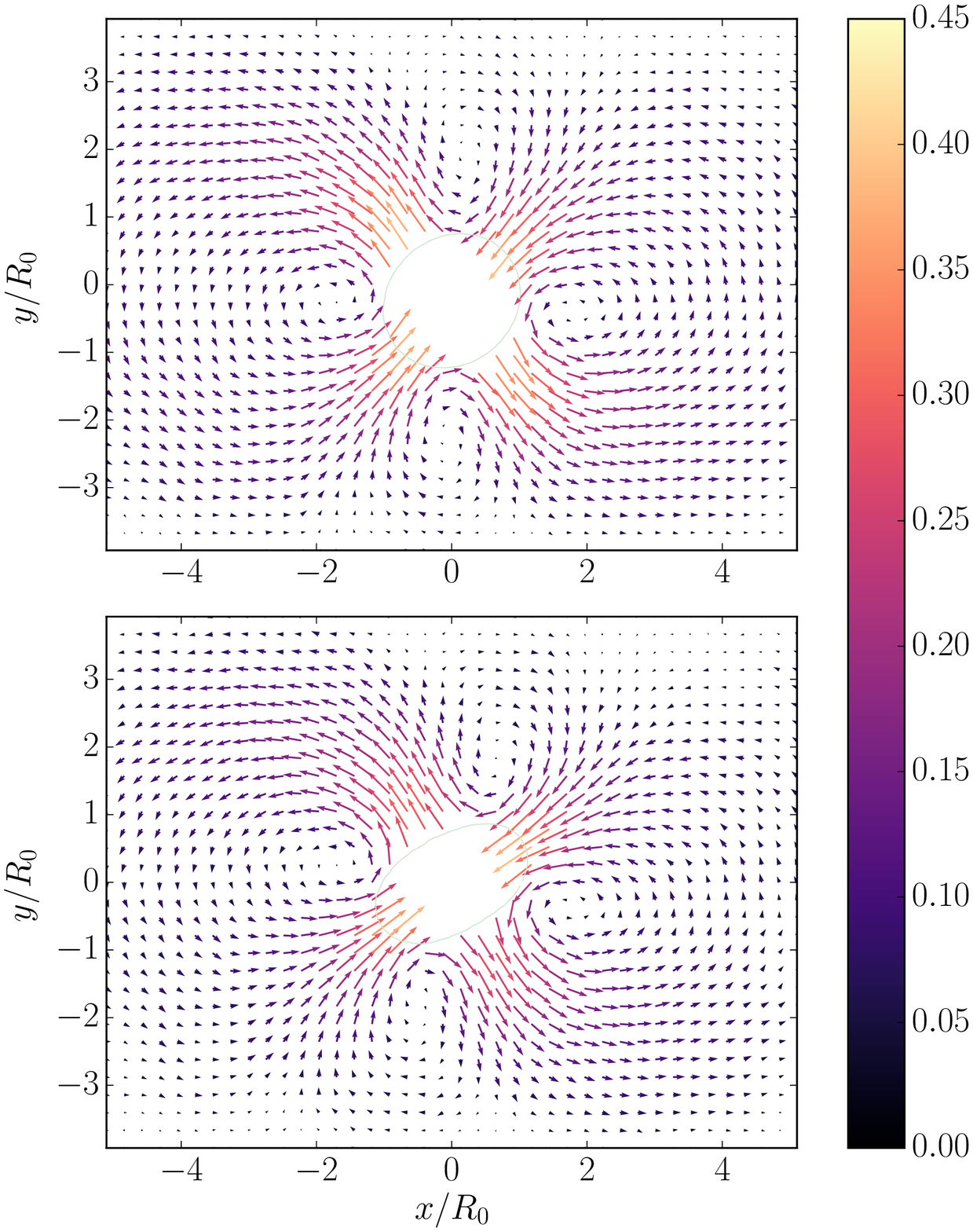}};
            \begin{scope}[x={(image.south east)},y={(image.north west)}]
              \draw (0.07,0.98) node {\ulcase{a}};
              \draw (0.07,0.49) node {\ulcase{c}};
              \draw (0.51,0.98) node {\ulcase{b}};
              \draw (0.51,0.49) node {\ulcase{d}};
            \end{scope}
  \end{tikzpicture}
  \caption{(Colour online) Time-averaged disturbance fields $\delta \vec{V}$: 
    \subcap{a} a quasi-spherical vesicle, experiment; 
    \subcap{b} a quasi-spherical vesicle with $A^*=0.984 \ (\Delta_L=0.05)$, numerical simulations; 
    \subcap{c} a deflated vesicle with $\Delta=0.39$, experiment; (d) a deflated
    vesicle with $A^*=0.885 \ (\Delta_L=0.40)$,
    numerical simulations. 
    Colours indicate the reduced disturbance magnitude $||\delta\vec{V}||/\dot{\gamma}R_0$. 
    The smooth contours indicates the vesicle shape and its mean position and
    orientation. 
    The line segments on the experimental data indicate the trajectory of the vesicle
    centre of mass; those included in the partial time averaging are marked by
    (orange) thicker lines in \subcap{c}.
    The plots of the experimental data are adapted from Figures 1\&2 of Ref.~\cite{Levant2012}.}
  \label{fig:velocityfieldexpsim}
  \end{center}
\end{figure*}

\begin{figure*}[tb]
  %\onefigure[width=16cm]{velocityfieldexpsim.eps}}
  \begin{center}
  \begin{tikzpicture}[font=\sffamily]%\bfseries\small]%Helvetika] 
        \node[anchor=south west,inner sep=0] (image) at (0,0)%
             {\includegraphics[width=0.47\textwidth,trim=0 6.5mm 10mm 2.5mm,clip]%
                                                        {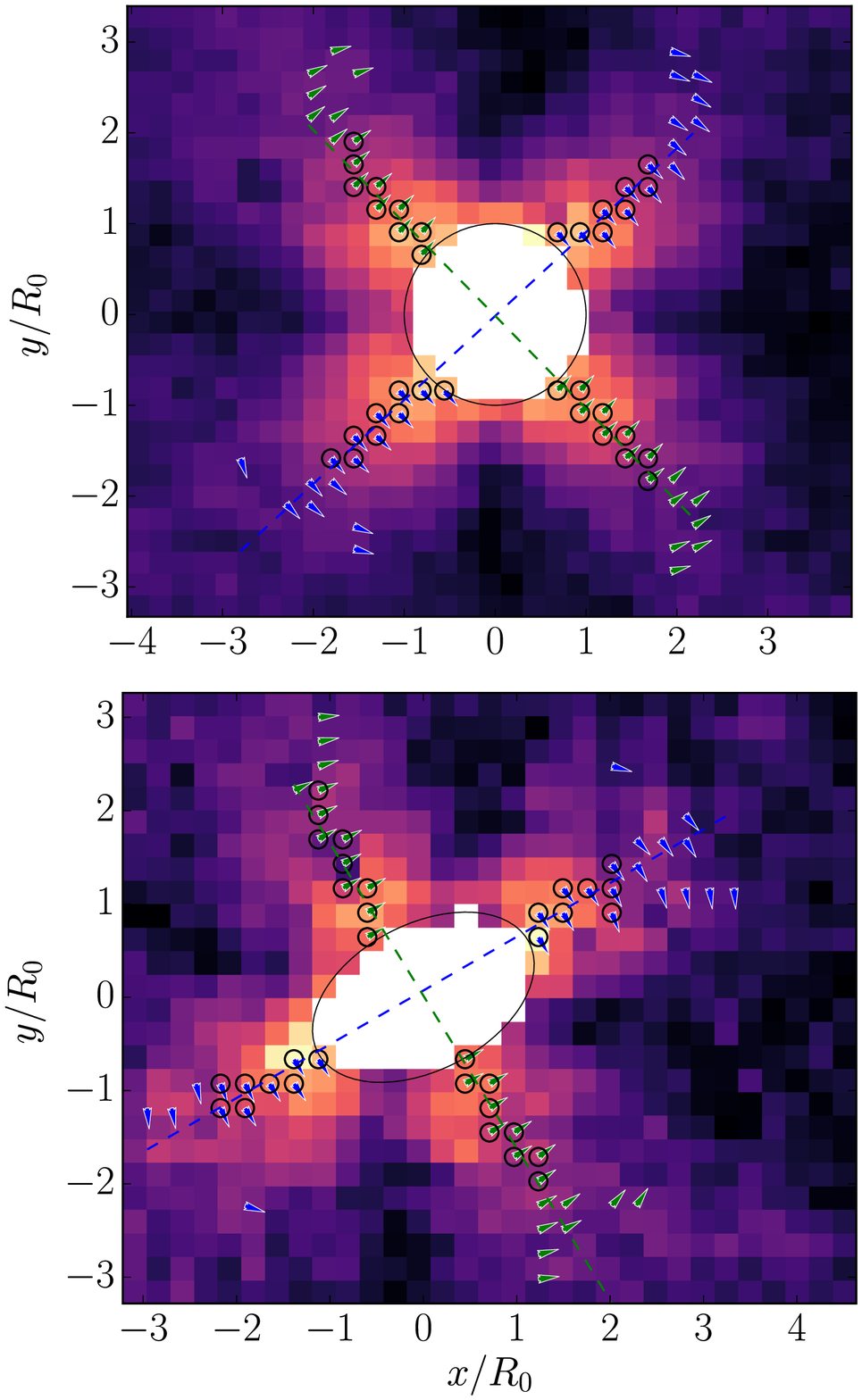}%
              \includegraphics[width=0.47\textwidth,trim=10mm 6.5mm 0 2.5mm,clip]%
                                                        {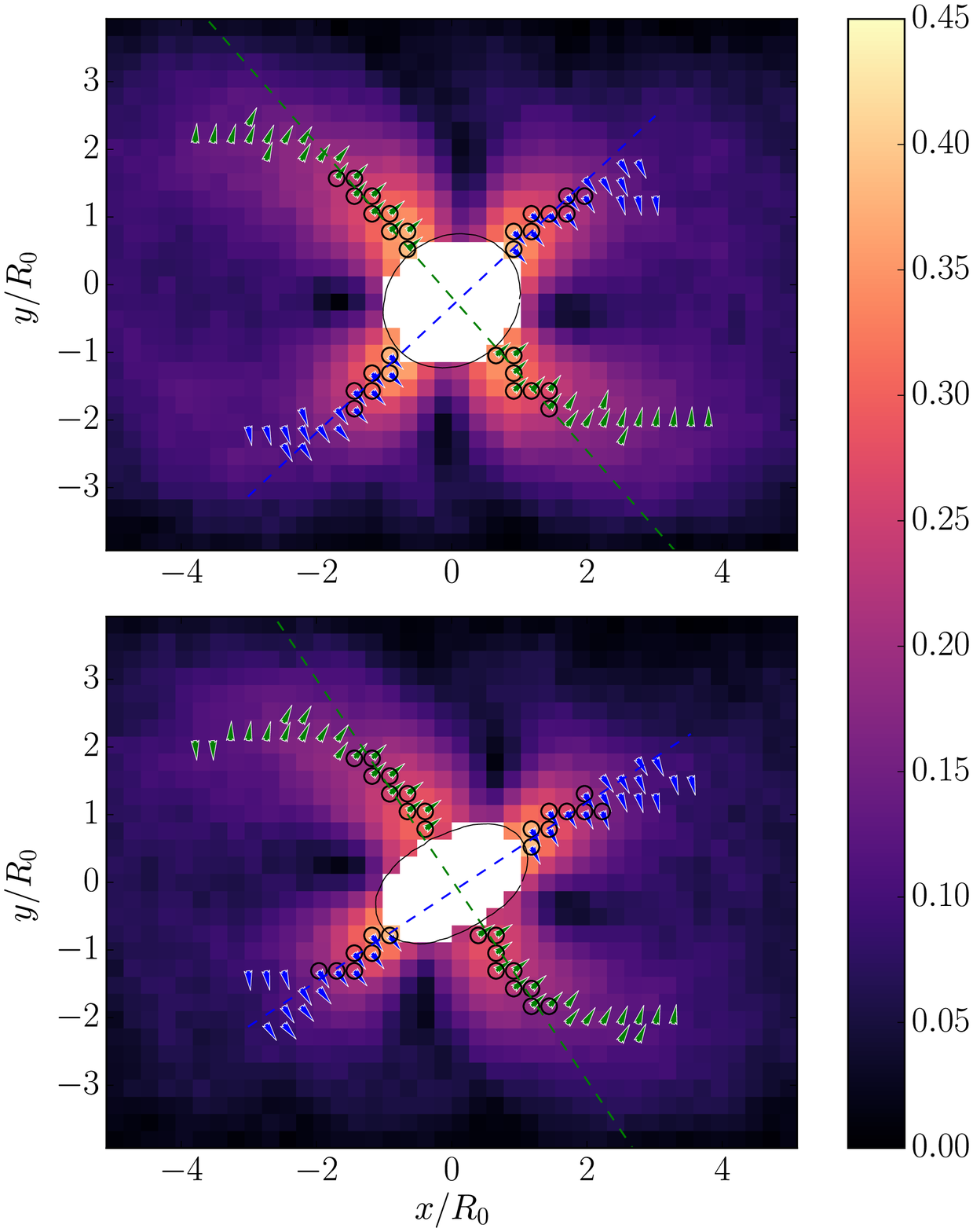}};
            \begin{scope}[x={(image.south east)},y={(image.north west)}]
              \draw (0.07,0.98) node {\ulcase{a}};
              \draw (0.07,0.49) node {\ulcase{c}};
              \draw (0.51,0.98) node {\ulcase{b}};
              \draw (0.51,0.49) node {\ulcase{d}};
            \end{scope}
  \end{tikzpicture}
  \caption{(Colour online) Time-averaged disturbance fields magnitude $\|\delta \vec{V} \|$: 
    \subcap{a} a quasi-spherical vesicle, experiment; 
    \subcap{b} a quasi-spherical vesicle with $A^*=0.984 \ (\Delta_L=0.05)$, numerical simulations; 
    \subcap{c} a deflated vesicle $\Delta=0.39$, experiment; 
    \subcap{d} a deflated vesicle $A^*=0.885 \ (\Delta_L=0.40)$, numerical simulations. 
    Colours indicate the reduced disturbance magnitude $||\delta\vec{V}||/\dot{\gamma}R_0$. 
    The black contour indicates the vesicle shape and orientation. 
    The arrow heads designate the detected ridges in this landscape of the
    disturbance magnitude; these are directed along the
    eigen-vector of the local Hessian matrix, corresponding the least principal
    curvature; blue and green colouring designate grouping by strain directions. 
    The circled ones reside within 2.5 reduced distance $d / R_0$
    from the vesicle centre of mass, and were included in the linear
    regression.
    The dashed lines show the best fitting linear regression results,
    approximating the ridges in the landscape of the disturbance magnitude
    $||\delta\vec{V}||$ (see details in the text).}
\label{fig:contourplots}
  \end{center}
\end{figure*}

After an initial transient, vesicles reach a tank-treading steady state while
diffusing, due to thermal noise, across the centre line of the channel.
Values for the reduced area $A^* = A_0/\pi R_0^2$, related to the excess length
$\Delta_L$ by $A^*=(1+\Delta_L/2\pi)^{-2}$, were chosen to be comparable with the
experimental ones: At the initial simulation time $t_0$ it is
$A^*(t_0) = 1$, corresponding to a nearly spherical vesicle, which in the flow settles
at $A^*=0.984$, and $A^*(t_0)
= 0.89$ for the deflated one, reaching $A^*=0.885$ in the steady flow.
The time averages of the measured vesicle inclination angles are
$\langle \theta \rangle = (0.226\pi, 0.162\pi$)~\si{\radian} with standard deviations 
$ \sigma_\theta = (0.048\pi, 0.025\pi)$~\si{\radian}, for $A^* = 0.984, 0.885$,
respectively.

A few notes related to the numerical conservation of vesicle length and area
are worth adding.
In the numerical model used here, the membrane is modelled
using beads successively connected by elastic bonds to form a closed ring (see
Ref. \cite{Finken2008} for details). The bond stiffness is the control parameter which
maintains the membrane length $L$ in proximity to its initial value $L_0$, 
both locally and globally. 
To respect the physical constraints, the vesicle area is kept close to the initial value
$A_0$ by introducing a quadratic potential depending on
a compression modulus \cite{Finken2008}.
Here, the deviations from the values $A_0$ and $L_0$ are below $0.6\%$ and $0.5\%$,
respectively. 
However, the values of bond stiffness and compression modulus 
cannot be arbitrarily increased  due to numerical instability.
This limitation is well known and appears in other membrane dynamics modelling,
erythrocytes for example \cite{Yazdani2011}, 
in particular at larger shear stresses.

As a quantitative measure for the strength of the vesicle effect on the
velocity field, we chose to study the decay of the disturbance field magnitude
$\| \delta \vec{V} \|$ with the distance from the vesicle.
To this end, the following analysis procedure was applied to both the experimental and the
numerical disturbance fields:
A ridge detection algorithm, explained in detail as part of Ref.
\cite{Afik2015a} (the code is available online), 
based on a differential geometric descriptor \cite{Lindeberg1999}, 
was applied to the $ \| \delta \vec{V} \|$ landscape. 
Position coordinates residing on a ridge were sorted based on their
location in the frame of reference whose origin is at the vesicle
centre of mass. Then, linear regression was applied to find the best fitting
lines approximating the principal ridge directions within a distance of up to 2.5 vesicle effective
radii.
To analyse $\| \delta \vec{V} \|$ along these directions, for every sampled $x$
coordinate within the data, a corresponding $y$ coordinate from the data grid
was chosen by minimal distance from the best fitting line.

\section{Results}

\begin{figure*}[tb]
  \newlength\scalingFigsWidth
  \setlength\scalingFigsWidth{.47\textwidth}
  \begin{center}
    \tikzLabelMod{a}{0.96}{%
       \includegraphics[width=\scalingFigsWidth,trim=0 11.34mm 0 5mm, clip]%
           {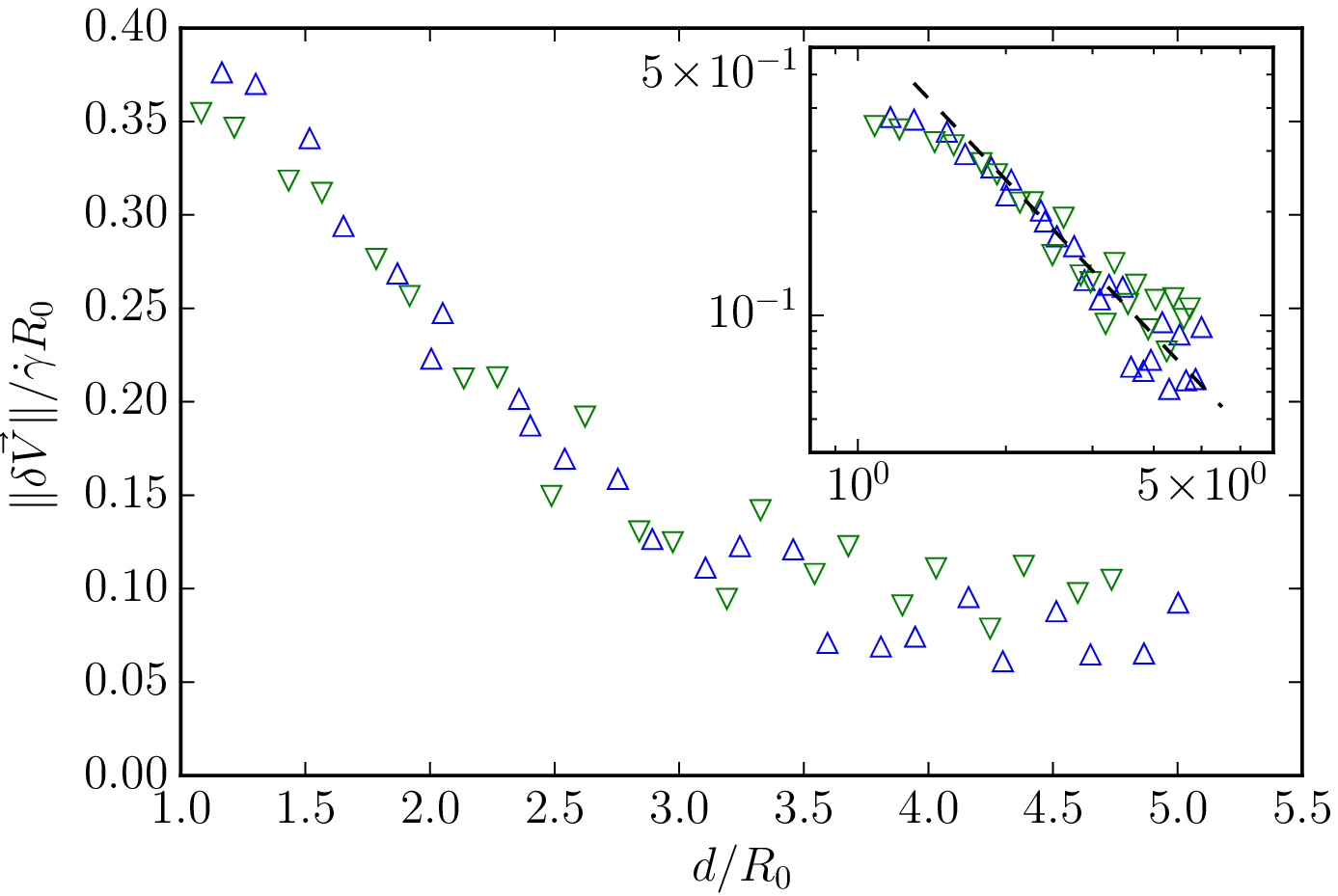}
           }%
    \tikzLabelMod{b}{0.96}{%
       \includegraphics[width=\scalingFigsWidth, trim=0 11.34mm 0 5mm, clip]%
           {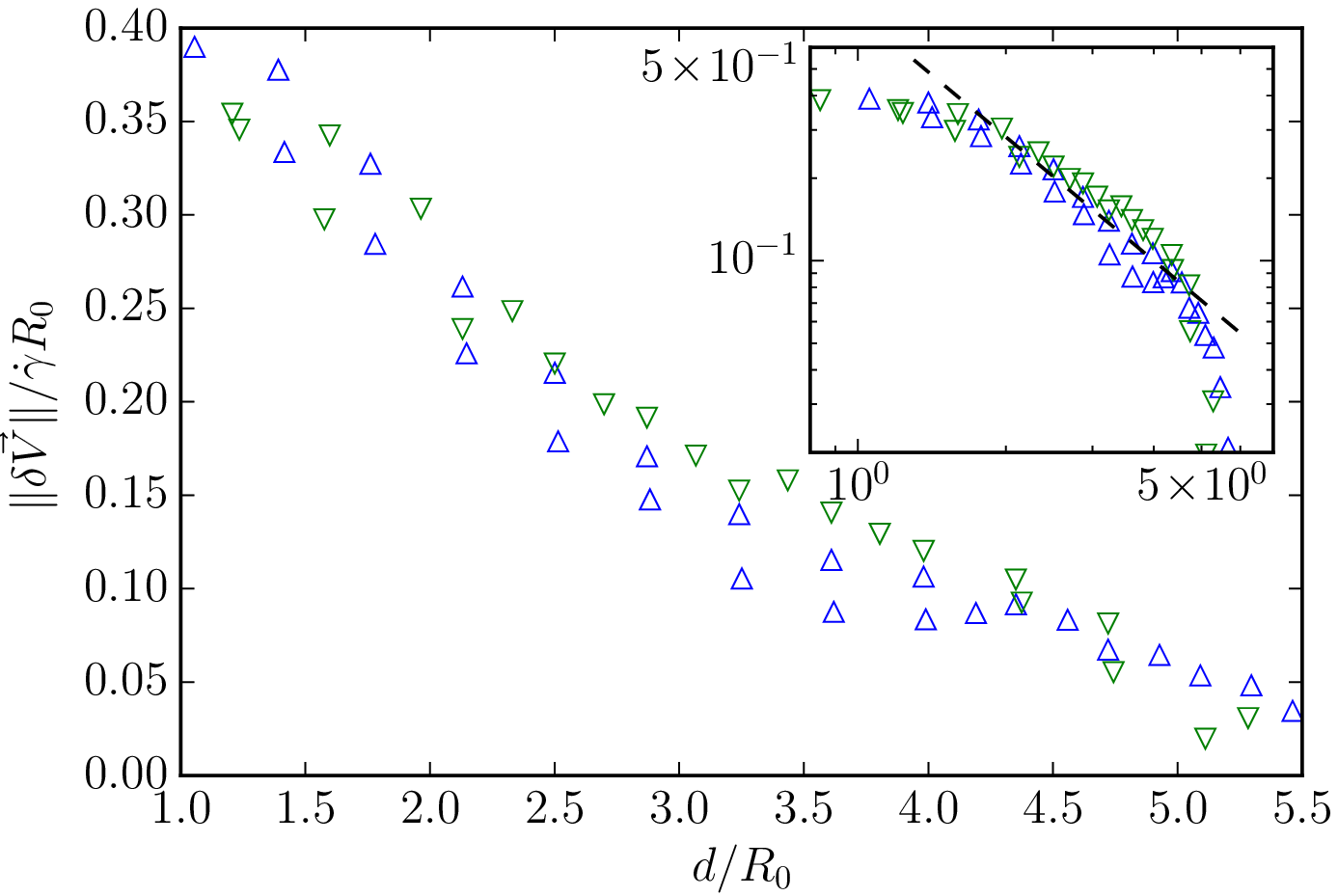}
           }
           
    \tikzLabelMod{c}{0.96}{%
       \includegraphics[width=\scalingFigsWidth,trim=0 0.5mm 0 5mm, clip]%
           {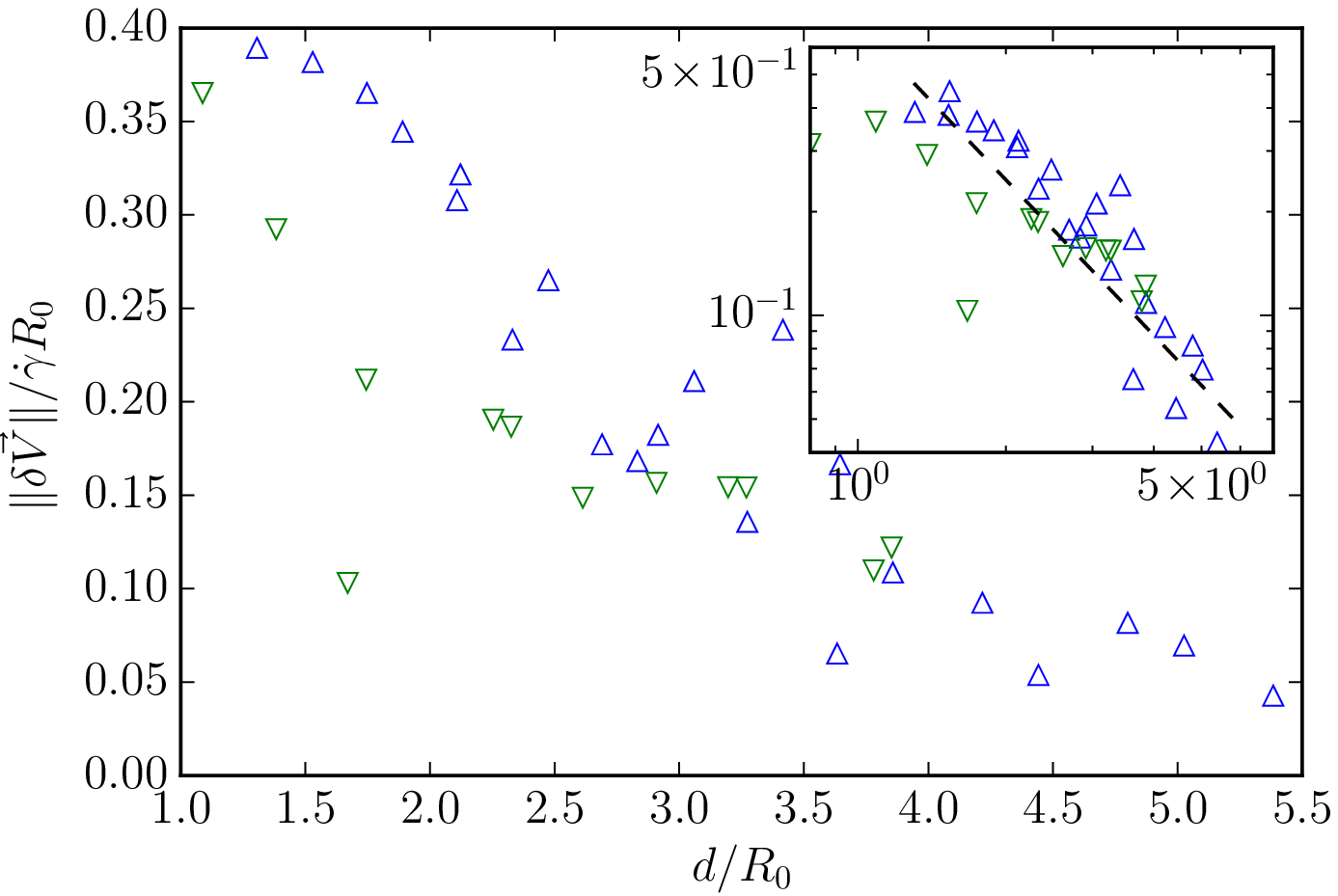}
           }%
    \tikzLabelMod{d}{0.96}{%
       \includegraphics[width=\scalingFigsWidth,trim=0 0.5mm 0 5mm, clip]%
           {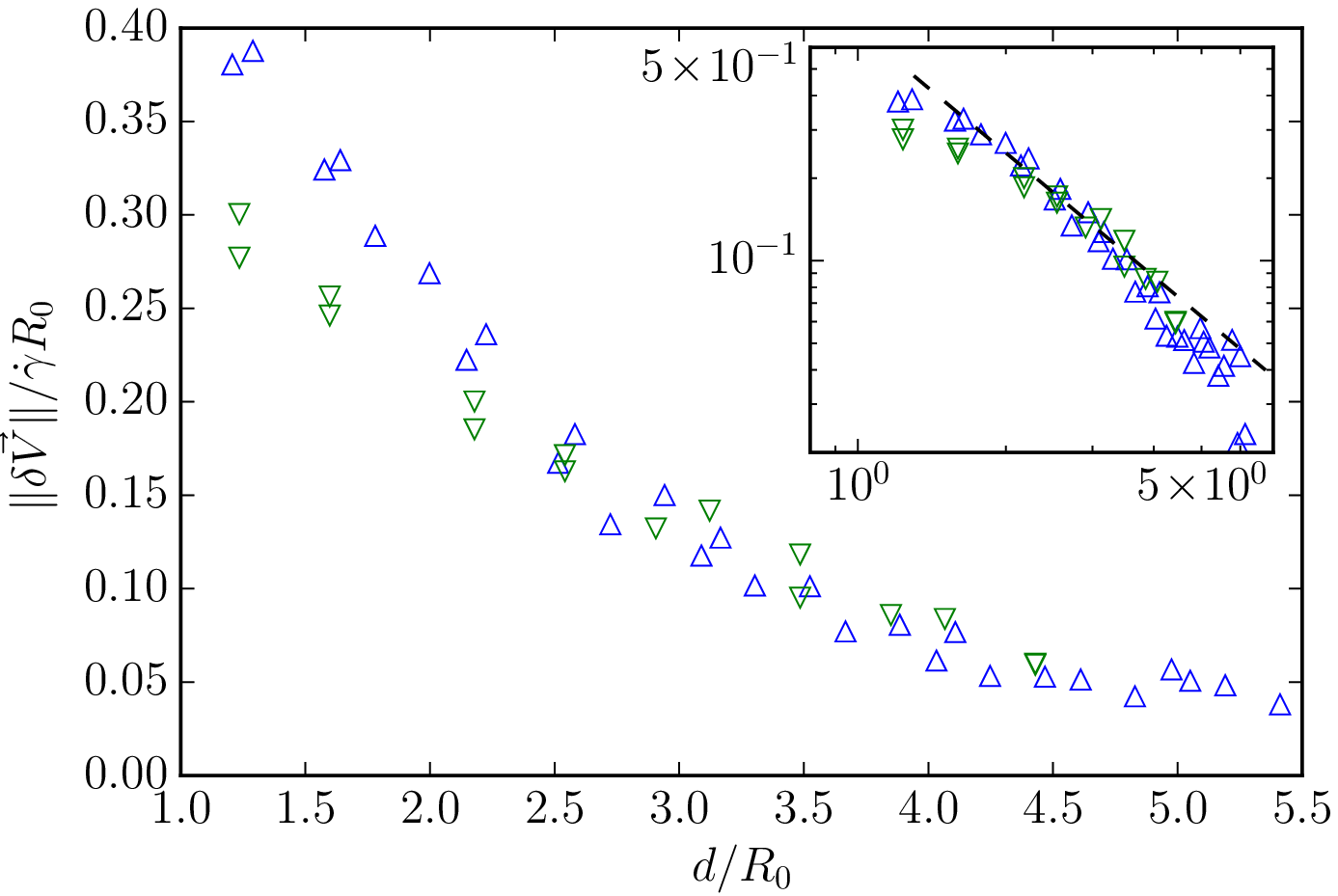}
           }
  \caption{(Colour online) The reduced disturbance field magnitude $||\delta\vec{V}||/\dot{\gamma}R_0$ 
           as a function of the reduced distance $d/R_0$ from the vesicle
           centre of mass in the vicinity of the regression lines; up/down
           triangles correspond to positive/negative slopes of the regression
           lines in Fig.~\ref{fig:contourplots}:
    \subcap{a} a quasi-spherical vesicle, experiment; 
    \subcap{b} a quasi-spherical vesicle $A^*=0.984  \ (\Delta_L=0.05$), numerical simulations.
    \subcap{c} a deflated vesicle $\Delta=0.39$, experiment; 
    \subcap{d} a deflated vesicle $A^*=0.885 \  (\Delta_L=0.40$), numerical
    simulations. 
    The insets: data presented on a log-log scale  (for clarity, axes range may be
    different from the linear axes). 
    The dashed lines indicate a power law decay $\propto (d/R_0)^{-3/2}$.
    }
    \label{fig:scalingspher}
    \end{center}
\end{figure*}

We characterise and quantify the effect of a single vesicle on the surrounding
velocity field which drives it, via the analysis of the disturbance field
$\delta \vec{V}$ it induces;
here $\delta \vec{V}$ is the difference between the velocity field in the absence and
presence of a vesicle.
Fig.~\ref{fig:velocityfieldexpsim} shows the resulting disturbance fields,
where the panels on the left show the experimental results and the right ones
show the numerical simulations; quasi-spherical vesicles are presented in the
top part, while the deflated ones can be found at the bottom. This
ordering is respected throughout Figs.~\ref{fig:velocityfieldexpsim}--\ref{fig:scalingspher}.

First we address the case of a quasi-spherical vesicle having a very small
excess surface area $\Delta$ (excess length $\Delta_L$ in the 2D numerical
simulations). In the experiments, the vesicle was located in the vicinity 
of the stagnation point of the dynamical trap and moving on a small orbit in a
planar linear flow of $\omega=0.46$ s$^{-1}$ and $\omega/s=1.8$ \cite{Levant2012}.
The second figure in Ref. \cite{Levant2012} shows the reference field (top panel), from
which the measured velocity field in the presence of a vesicle (middle panel) 
was subtracted, resulting in the disturbance field (bottom panel); 
an adapted plot of the disturbance field is provided here in
Fig.~\ref{fig:velocityfieldexpsim}\subfig{a}. 
It allows to study the back-reaction of the
vesicle on the flow and to estimate the effective distance of this
back-reaction. Note that the characteristic disturbance field magnitude is
smaller than the velocities in the reference field by more than an order of
magnitude (see Fig.~2 in \cite{Levant2012}). Two pronounced qualitative
features of the disturbance velocity field are evident already at this stage:
\begin{inparaenum}[(a)]
  \item the flipping of strain directions in the vicinity of the vesicle, which
    is significant up to a few vesicle effective radii along the strain main
    axes, and 
  \item the presence of four vortices generated in the vicinity of the vesicle%
\end{inparaenum}.
The former can be viewed as a result of the tank-treading motion of the
membrane, which moves at a uniform tangential velocity throughout the membrane, thus transferring
momentum from the shear direction to the perpendicular one. The uniformity of
the membrane tangential velocity is due to the vesicle's volume and area
conservation, which opposes extension and compression.
These main features of the vesicle back-reaction on a shear flow are captured
in the 2D numerical simulations, as can be seen from  the time-average
of the disturbance field around a nearly spherical vesicle provided in
Fig.~\ref{fig:velocityfieldexpsim}\subfig{b}.
They were also found and presented in a snapshot of the velocity field around a
nearly spherical vesicle in Fig.~1 of Ref.~\cite{Finken2008}, as well as in snapshots
of the disturbance field generated by a solid spherical particle in a
shear flow obtained in 3D numerical simulations \cite{Sangani2011}.
It seems that dimensionality does not play a key role in determining the flow
features of interest to us in this work.
A similar disturbance field and streamlines caused by a torque-free solid sphere
in a simple shear flow were presented in Fig.~9 of Ref.~\cite{Mikulencak2004}, where
the results of numerical simulations for $Re=0,1,10$ were shown. However, one
notices that in the $Re=0$ case the streamlines are straight and the
vortices are absent \cite{Mikulencak2004}. 
At $Re=1$ the streamlines are slightly curved and two
vortices in the compressed direction become visible in sharp contrast to
the perturbed velocity field caused by a solid spherical particle in
Ref.~\cite{Sangani2011} and by a spherical vesicle at $Re \ll 1$ in both the
experimental results and numerical simulations reported here 
(see  Fig.~\ref{fig:velocityfieldexpsim}\subfig{a},\subfig{b}), 
where four vortices are evident.

Next, we proceed with the experimental results of a deflated vesicle moving on a
small orbit in a plane linear flow with $\omega=0.46$ s$^{-1}$ and
$\omega/s=1.4$; due to the comparable size of the vesicle and its orbit, we
performed partial time averaging including only the sections highlighted by
(orange) thicker lines in Fig.~\ref{fig:velocityfieldexpsim}\subfig{c}, adapted from Fig.~3 of
Ref.~\cite{Levant2012}.
The corresponding numerical results are presented in Fig.~\ref{fig:velocityfieldexpsim}\subfig{d}. 
These are qualitatively consistent with the experimental ones,
Fig.~\ref{fig:velocityfieldexpsim}\subfig{c}.
A quantitative comparison to the nearly spherical case, as well as between the
experimental and the numerical results, can be found in what follows.

In order to quantitatively analyse the disturbance field, we focus on
the its magnitude landscape $\| \delta \vec{V} \|$, plotted in Fig.~\ref{fig:contourplots} as
heat maps.
The main features of these landscapes were extracted using the application
of a ridge detection algorithm \cite{Afik2015a}, as explained in the previous
section. 
The arrow heads in Fig.~\ref{fig:contourplots} designate the detected ridge loci, pointing
perpendicular to the ridge (they are directed along the eigen-vectors of the
local Hessian matrix, corresponding to the least principal curvature); 
colours correspond to the grouping by stretching/compression directions. 

We next wanted to study the functional dependence of the disturbance magnitude
decay on the distance from a vesicle. The principal directions of the ridges
were extracted using a linear approximation at the vicinity of the vesicle
within the dimensionless distance from the vesicle center of mass
of $d / R_0 \lesssim 2.5$ (corresponding
ridge loci are circled in Fig.~\ref{fig:contourplots}, the linear regression
results are plotted as dashed lines), and extrapolated based
on these best fitting lines farther on up to $d/R_0\approx 6$. 
Based on the experimental data for the vesicles at $\lambda=1$, the angles
formed by the approximating lines with respect to the flow direction 
are $\varphi_\pm \approx ( 0.237\pi , - 0.254\pi)$~\si{\radian} in the nearly spherical case,
and $\varphi_\pm \approx (0.166\pi, -0.325\pi)$~\si{\radian} in the deflated case.
The pronounced difference between the deflated vesicle and the nearly spherical
one is in the deviation of the angles  $\varphi_\pm$ from the strain principal directions: In the case of the deflated vesicle we
study here, the stretching direction deviates from the strain direction by
about $\pi/12$~\si{\radian}. 
\iffalse
the experimental deflated vesicle is at <\theta> \approx 27.65 degrees 
\fi
The angles resulting from the analysis of the numerical data 
are $\varphi_\pm \approx (+0.239\pi , - 0.270\pi)$~\si{\radian}
and $\varphi_\pm \approx (+0.186\pi, -0.311\pi)$~\si{\radian} 
for reduced areas $A^*=0.984$ and $0.885$, respectively.
Note that the principal directions are nearly orthogonal to each other, that is
$\varphi_+ -\varphi_- \approx \pi/2$~\si{\radian}, and that the positive angle
$\varphi_+$ is consistent with the average inclination angle 
$\langle \theta \rangle$ for both values of the reduced area.

The values of the reduced disturbance field magnitude $\| \delta \vec{V} \| /\dot{\gamma}R_0$
in the vicinity of the regression lines are plotted in Fig.~\ref{fig:scalingspher}
as a function of the reduced distance $d/R_0$ from the vesicle centre of mass. 
The insets in Fig.~\ref{fig:scalingspher} present these data sets on a log-log
scale, indicating that there is an intermediate regime where the decay can be
approximated by a power-law scaling $\propto (d/R_0)^{-3/2}$ to a reasonable
degree. 
The experimental data for the deflated vesicle exhibit a larger scatter, yet
show no contradiction to the power law decay with the same exponent.
This larger scatter is attributed mainly to the averaging over a
smaller data set, with respect to one of the quasi-spherical vesicle. In the
deflated case, the vesicle orbit was comparable to its size, hence averaging
was taken over a sub-sample where the vesicle spent the larger fraction of
time.

\begin{figure}[t]
   \centering
   \includegraphics[width=\columnwidth,trim=0 0 8mm 6mm, clip]%
        {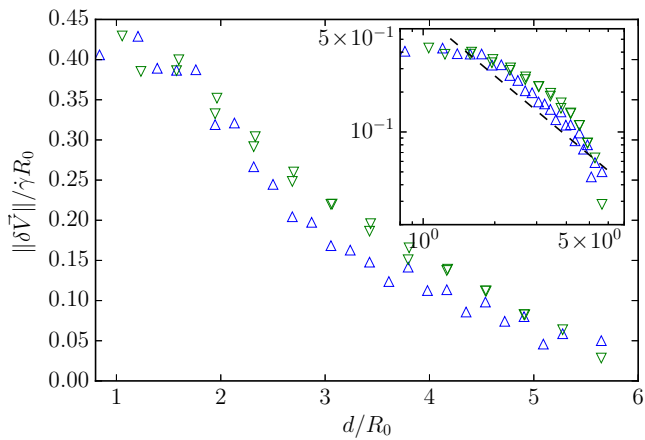}
   \caption{(Colour online) The reduced disturbance field magnitude
     $||\delta\vec{V}||/\dot{\gamma}R_0$ as a function of the reduced distance $d/R_0$
     from the vesicle centre of mass of a quasi-spherical vesicle
     $A^*=0.986 \ (\Delta_L=0.04)$ at viscosity contrast $\lambda=20$, numerical simulations. 
     Up/down triangles correspond to positive/negative slopes of regression lines.
     The inset: data presented on a log-log scale. 
     The dashed line corresponds to power-law decay $\propto (d/R_0)^{-3/2}$.}
   \label{fig:scalingspher20}
  \end{figure}

Finally, we wanted to assert that numerical approach employed in this work
captures the expected spherical solid-body behaviour at high enough values of viscosity contrast.
Analysing the numerical results of a quasi-spherical vesicle at
$\lambda=20$ resulted in the following angles for the principal directions
$\varphi_\pm \approx (+0.250\pi , - 0.256\pi)$~\si{\radian}.
Indeed, this is close to what is expected for a solid sphere.  
The disturbance magnitude decay with distance, presented in
Fig.~\ref{fig:scalingspher20} for comparison, exhibits a scaling
behaviour similar to the one previously described.

\section{Discussion and Conclusions}

In this work we have studied the back-reaction of a single vesicle,
undergoing tank-treading motion, on the driving planar linear velocity field which surrounds
it. 
Here we focused on the disturbance field induced by the vesicle, which results
from the subtraction of the velocity field when the vesicle is present, from the reference
field without it.
The approach presented was applied to experimental results as well as to
numerical simulations at comparable vesicle and flow configurations.
The comparison between the wet experiment and the computerised one reveals an
impressive agreement.

The landscape of the disturbance field magnitude was analysed to achieve
quantitative results. 
We tracked ridges in this landscape and found two principal directions of
vesicle effect, opposing the strain compression and stretching directions.
The two principal directions are nearly perpendicular to each other and
approximately follow the average tilt angle of the vesicle.

The decay of the disturbance field magnitude with the distance $d$ from the vesicle
along these principal directions reveals an intermediate regime which can be characterised as
a power law decay $\|\delta \vec{V} \| /\dot{\gamma}R_0 \propto (d/R_0)^{-3/2}$.
This result seems to be supported by our data for a nearly spherical vesicle, as well as
for a moderately deflated one.
The reported scaling for the decay is different from the theoretical prediction for long-range
hydrodynamic effect at large distances $\sim 1/d +1/d^2 +O(1/d^3)$ at $d/R_0 \gg 1$,
resulting from a membrane in an external shear flow and obtained as a special solution of the inhomogeneous
Stokes equation for the velocity field \cite{Happel1983}.

The effect of a single vesicle undergoing tank-treading motion in a planar linear flow remains
significant up to a distance of about 4 effective vesicle radii. 
This value is consistent with the separation distance between two tank-treading
vesicles in the same flow, at which their dynamics is altered
\cite{Levant2012}.
Unfortunately, an attempt to reproduce numerically the experimental result on
the dependence of the root-mean-squared fluctuations of the deflated vesicle 
inclination angle as a function of distance from a spherical one, was not
successful so far and resulted in very scattered data. Nevertheless, the
results we found in numerical simulations showed no contradiction to the
scaling found experimentally \cite{Levant2012}.

To summarise, we found an indication for the long-range hydrodynamic effects of a
single vesicle on the velocity field, which are indeed present in a planar linear
flow; this conclusion is supported both by experimental results and numerical simulations.
Therefore, this calls for the incorporation of such long-range hydrodynamic
effects into future models for improved predictions of the rheology of a vesicle suspension.

 \acknowledgments 
This work was partially supported by grants from Israel Science Foundation and
Volkswagen Foundation through the Lower Saxony Ministry of Science and Culture
Cooperation (Germany).

%\bibliographystyle{eplbib}
%\bibliography{EA_academic_references}
%% TODO include inline bbl file removing \newline\url{ ... }
%\input{Long-rangehydrodynamiceffect.bbl}

\end{document}